\documentstyle[prl,aps,preprint,floats]{revtex}

\input BoxedEPS
\SetRokickiEPSFSpecial  
\def\half{\frac{1}{2}}
\HideDisplacementBoxes
\tightenlines
\begin{document}

\title{Fermionic One-Loop Corrections to Soliton Energies in 1+1~Dimensions}

\author{N.~Graham\footnote{graham@mitlns.mit.edu}
and R.~L.~Jaffe\footnote{jaffe@mit.edu}}

\address{{~}\\Center for Theoretical Physics, Laboratory for
  Nuclear Science
  and Department of Physics \\
  Massachusetts Institute of Technology,
  Cambridge, Massachusetts 02139 \\
  {\rm MIT-CTP\#2813 \qquad hep-th/yymmddd}}

\maketitle

\begin{abstract}
We demonstrate an unambiguous and robust method for computing
fermionic corrections to the energies of classical background field
configurations.  We consider the particular case of a sequence of
background field configurations that interpolates continuously
between the trivial vacuum and a widely separated soliton/antisoliton
pair in 1+1~dimensions.  Working in the continuum, we use phase
shifts, the Born approximation, and Levinson's theorem to avoid
ambiguities of renormalization procedure and boundary conditions.  We
carry out the calculation analytically at both ends of the
interpolation and numerically in between, and show how the relevant
physical quantities vary continuously.  In the process, we elucidate
properties of the fermionic phase shifts and zero modes.
\end{abstract}

 \narrowtext

\section{Introduction}

1+1~dimensional models have pointed out many subtleties in the
computation of one-loop fermionic quantum corrections to the energies
of classical field configurations.  Many works, usually in the
context of supersymmetric theories \cite{many}, have reached
conflicting conclusions about the correct boundary conditions,
regularization and renormalization procedures, and bound state and
continuum contributions.  Since the one-loop
energy is given by the cancellation of a divergent sum of
zero-point energies against a divergent counterterm, one must be
extremely careful to fix the counterterm with a definite
renormalization scheme in order to obtain a physically meaningful
result.

In this paper we extend the methods of Ref.~\cite{us2,us4} to this case.
We again find that 1+1~dimensional models provide an excellent testing
ground for our methods, and can give considerable insight into the
causes and resolutions of the disagreements among the previous results.
As in our previous works, we will use a
continuum formalism based on phase shifts, their Born approximations,
and Levinson's theorem.  Relying on such physical quantities results
in a procedure that is both analytically unambiguous and numerically
robust.  As a result, we will be able to compare numerical results in
generic, but not exactly soluble cases to analytic results in
specific, exactly soluble cases.

We will consider a Majorana fermion coupled to a real scalar
background field, with a definite choice of bosonic and fermionic
potentials, though our method applies equally well for any other
choice of potentials, as well as for complex scalars and Dirac
fermions (including cases without $C$ or $CP$ invariance).  We
will consider a continuous sequence of scalar field configurations
that interpolates between a trivial background and a widely separated
soliton/antisoliton pair, always with the same, trivial field
configuration as $x\to\pm\infty$.  In this way we can unambiguously
track physical quantities
such as phase shifts and bound state energies, staying in the trivial
sector of the theory throughout the process.  In so doing, we obtain
results numerically that continuously approach the results obtained
analytically in Ref.~\cite{us}.

In \S II we develop the methods we will need to compute
phase shifts, which in turn allow us to compute the renormalized
one-loop energy as a sum over zero-point energies $-\half\omega$.  In
\S III we apply these methods to the specific sequence of field
configurations we have chosen and obtain numerical and analytical results.
In \S IV we summarize our results, and connect them to previous and
future works.  The Appendix contains techniques parallel to those of
\S II but for a single soliton instead of a soliton/antisoliton pair.

\section{Fermionic small oscillations and phase shifts}

We consider a Majorana fermion $\Psi$ interacting with a scalar background
field $\phi$, with the classical Lagrangian density
\begin{equation}
  {\cal L} = \frac{m^2}{2\lambda} \left(
    i\bar\Psi \slash\hspace{-0.5em}\partial \Psi - m\phi \bar \Psi
    \Psi + {\cal L}_\phi \right)
  \label{LaG}
\end{equation}
where ${\cal L}_\phi$ is the Lagrangian density for the $\phi$
background field.  As discussed in the Introduction, we can treat
any bosonic potential.  To be specific, we choose
\begin{equation}
{\cal L}_\phi = \half (\partial_\mu \phi)(\partial^\mu \phi) -
\frac{m^2}{8}(\phi^2-1)^2
\label{Lphi}
\end{equation}
which has classical soliton (antisoliton) solutions satisfying
\begin{equation}
\phi^\prime(x) = \mp \frac{m}{2}(\phi^2 - 1)
\end{equation}
giving
\begin{equation}
\phi(x) = \pm \tanh \frac{mx}{2}.
\end{equation}
For the most part, we will be treating $\phi$ as a classical
background, so that its dynamics will be unimportant.  However, we
note that choosing the Lagrangian density of Eq.~(\ref{Lphi}) causes the
bosonic and fermionic degrees of freedom to be related by
supersymmetry for a background field that is a solution to the
equations of motion (such as a single soliton or an infinitely
separated soliton/antisoliton pair).  In what follows, we will find it
instructive to compare the bosonic and fermionic small oscillations
spectra.  The supersymmetric model is discussed in more detail in 
Ref.~\cite{us}.

The one-loop corrections to the energy due to fermionic fluctuations
will be given by the appropriately renormalized sum of the zero-point
energies, $-\half\omega$, of the fermionic small oscillations.  The
spectrum of fermionic small oscillations in a background $\phi_0(x)$
is given by the Dirac equation,
\begin{equation}
  \gamma^0 \left(-i\gamma^1 \frac{d}{dx} + V_F(x)\right) \psi_k(x) =
\omega^F_k
  \psi_k(x)
  \label{Diraceq}
\end{equation}
where $V_F(x) = m\phi_0(x)$ is the fermionic potential and
$k=\pm\sqrt{\omega^{2}-m^{2}}$ is the momentum labeling the scattering
states.

We will choose the convention $\gamma^0 = \sigma_2$ and $\gamma^1 = i
\sigma_3$ so that the Majorana condition becomes simply $\Psi^\ast = \Psi$.
We note that since our Lagrangian is $CP$ invariant, all of our
results for the spectrum of a Majorana fermion can be extended to a
Dirac fermion simply by doubling.

Our primary tool for analyzing the small oscillations will be the
phase shifts.  We can solve for the phase shifts of any fermionic
potential $V_F(x)$ that satisfies $V_F(x) = V_F(-x)$ and $V_F(x)\to m$
as $x\to\pm\infty$ by generalizing the
methods of Ref.~\cite{us2} to fermions.
This form will be useful for considering our example of a sequence of
background field configurations that continuously interpolates between
the trivial vacuum and a widely separated soliton-antisoliton pair.
In the limit of infinite separation, the phase shift for the pair goes
to twice the result for a single soliton.  For comparison, we also do
the computation for a single soliton directly in the Appendix.

We define the parity operator acting on fermionic states as $P=\Pi
\gamma^0$, where $\Pi$ sends $x\to -x$.  $P$ commutes with the
Hamiltonian, so parity is a good quantum number.  As a result, we can
separate the small oscillations into positive and negative parity
channels, now restricted to $k>0$ in the continuum.  

We parameterize the fermion wavefunction by
\begin{equation}
\psi(x) = \left( \matrix{
        e^{i\nu(x)} \cr
        i e^{i\zeta(x)} e^{i\theta}
} \right) e^{ikx}
\end{equation}
where $\theta = \tan^{-1} \frac{k}{m}$ and $\nu$ and $\zeta$ are
complex functions of $x$.  We then find the phase shift in each
channel $\delta^{\pm}(k)$ by solving Eq.~(\ref{Diraceq}) subject to
the boundary conditions
\begin{equation}
\psi^{\pm}(0) \propto \left( \matrix{ 1 \cr \pm i } \right).
\end{equation}
We obtain
\begin{eqnarray}
\delta^+(k) &=& -{\rm Re} \: \nu(0) + \frac{\theta}{2} + \frac{1}{2i}
\log \frac{Y-1}{1-Y^\ast} \cr
\delta^-(k) &=& -{\rm Re} \: \nu(0) + \frac{\theta}{2} + \frac{1}{2i}
\log \frac{1+Y}{1+Y^\ast}
\end{eqnarray}
where $Y = \frac{1}{\omega}(V_F(0) - ik + i\nu^\prime(0)^\ast)$
and $\nu(x)$ satisfies
\begin{equation}
-i\nu(x)^{\prime\prime} + \nu^\prime(x)^2 + 2k\nu^\prime(x) +
V_F(x)^2 - V_F(x)^\prime - m^2 = 0
\label{nueq}
\end{equation}
with the boundary condition that $\nu(x)$ and $\nu^\prime(x)$ vanish
at infinity.  The total phase shift is given by summing the phase shifts
in each channel, $\delta_F(k) = \delta^+(k) + \delta^-(k)$.

Our calculation  also requires us to find the
bound state energies, which we do by solving
Eq.~(\ref{Diraceq}) using  ordinary ``shooting'' methods.  However,
once we know the phase shifts, Levinson's theorem tells us how many
bound states there will be.  It works exactly the same way as in the
bosonic case \cite{us,lev}: In the odd parity channel, the number of
bound states $n_-$ is given by
\begin{equation}
\delta^-(0) = \pi n_-
\label{Lev1}
\end{equation}
while in the even parity channel the number of bound states $n_+$ is given
by
\begin{equation}
\delta^+(0) = \pi (n_+ - \half).
\label{Lev2}
\end{equation}
One can derive this result either by the same Jost function
methods used in the boson case, or by observing that at small $k$, the
nonrelativistic approximation becomes valid so the bosonic results carry
over directly.  For a particular potential there may exist a $k=0$ state in
either of the two channels whose Dirac wavefunction goes to a constant
spinor as $x\to\pm\infty$. (Generically, for $k=0$ the components of the
Dirac wavefunction go to straight lines as $x\to\pm\infty$, but not lines
with zero slope.)  Just as in the bosonic case, such threshold states
(which we term ``half bound states'') should be counted with a factor of
$\half$ in Levinson's theorem.

Given the phase shifts and bound state energies, we can calculate the 
one-loop fermionic correction to the energy.  We work in a simple 
renormalization scheme in which we add a counterterm proportional to
$\phi^2 - 1$, and perform no further renormalizations.  The counterterm is
fixed by requiring that the tadpole graph cancel.  We must sum the 
zero-point energies $-\half \omega$ of the small oscillations,  subtract
the same sum for the free case, and add the contribution of  the
counterterm.  As in Ref.~\cite{us2,us4,us}, we use the density of  states
\begin{equation}
\rho(k) = \rho_0(k) + \frac{1}{\pi} \frac{d\delta_F(k)}{dk}
\end{equation}
to write
\begin{equation}
\Delta H = -\half \sum_j \omega_j - \int_0^\infty \frac{dk}{2\pi}
\omega \frac{d\delta_F(k)}{dk}  + \Delta H_{\rm ct}
\end{equation}
where $\omega = \sqrt{k^2+m^2}$  and the sum over $j$ counts bound states
with appropriate factors of $\half$ as discussed above.  Next, we use
Eq.~(\ref{Lev1}) and Eq.~(\ref{Lev2}) to rewrite this expression as
\begin{equation}
\Delta H = -\half \sum_j (\omega_j-m) - \int_0^\infty \frac{dk}{2\pi}
(\omega-m) \frac{d\delta_F(k)}{dk}  + \Delta H_{\rm ct}.
\end{equation}
The (divergent) contribution from the tadpole graph is given by the leading
Born approximation to $\delta_F$.  Since we have chosen the counterterm
to exactly cancel the tadpole graph, we can rewrite the counterterm
contribution in terms of the Born approximation, giving
\begin{equation}
\Delta H = -\half \sum_j (\omega_j -m) - \int_0^\infty \frac{dk}{2\pi}
(\omega - m) \frac{d}{dk} (\delta_F(k) - \delta_F^{(1)}(k))
\label{totenergy}
\end{equation}
where the subtraction $\delta_F^{(1)}(k)$ consists of the first Born
approximation to $\delta_F(k)$ plus the part of the second Born approximation
related by $\phi\to-\phi$ symmetry,
\begin{equation}
\delta^{(1)}(k) = -\frac{1}{k} \int \left(V_F(x)^2 -  m^2 \right) \, dx
\end{equation}
which can also be obtained numerically by iterating
Eq.~(\ref{nueq}). As expected, it is indeed proportional to $\phi^2-1$.

\section{Numerical Results}

We study a sequence of background fields labeled by a parameter $x_0$ that
continuously interpolates from the trivial vacuum $\phi(x)=1$
at $x_{0}=0$ to a widely separated soliton/antisoliton
pair at $x_{0}\to\infty$,
\begin{equation}
\phi(x,x_0) = 1 + \tanh \frac{m(x-x_0)}{2} - \tanh \frac{m(x+x_0)}{2}.
\end{equation}
\begin{figure}[htbp]
$$
\BoxedEPSF{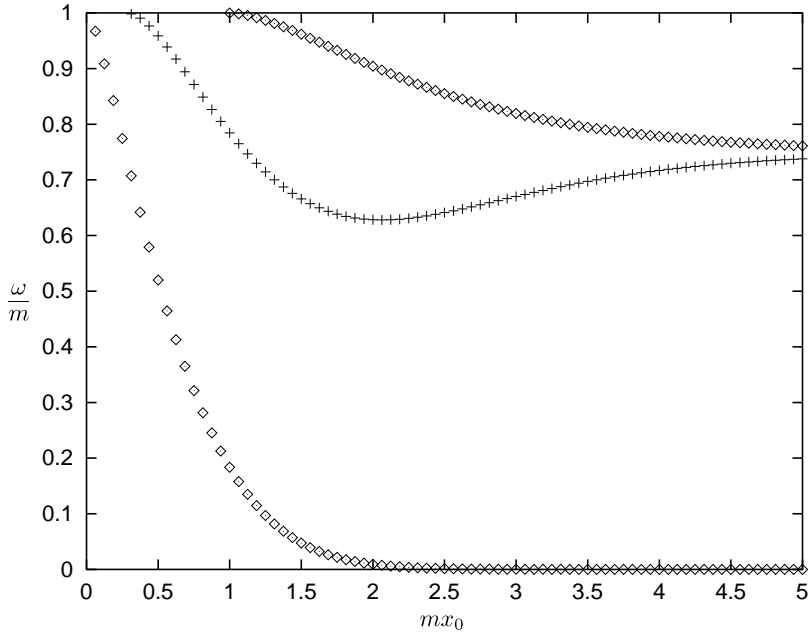 scaled 600}  
$$
\caption{Fermion bound state energies as a function of $mx_0$ in units
of $m$.  Even parity states are denoted with $\diamondsuit
\hspace{-0.525em} \cdot$~~and odd parity states with $+$.}
\label{boundf}
\end{figure}

Fig.~\ref{boundf} shows the fermionic bound state energies as a
function of $x_0$.  In the limit $x_0\to\infty$ two bound states approach
energy $m \frac{\sqrt{3}}{2}$.  These are  simply
the odd and even parity combinations of the single soliton
bound state at $m \frac{\sqrt{3}}{2}$.
The third (positive parity) bound state approaches $\omega = 0$
where the single soliton also has a bound state, but there
is only {\it one} such state (with $\omega >0$).  This
is consistent with the result obtained analytically
in Ref.~\cite{us}.  As emphasized in Ref.~\cite{us}, for a single
soliton we must thus count the zero mode with a factor of $\half$.

For any finite $x_{0}$, Levinson's theorem holds
without subtleties; there  are no states that require factors of
$\half$.  Thus for a large but finite $x_{0}$, we find
\begin{eqnarray}
\delta^+(0) &=& \frac{3\pi}{2} \cr \delta^-(0) &=& \pi
\end{eqnarray}
consistent with having two positive parity and one negative parity
bound states (see Fig.~\ref{boundf}).  In the limit $x_0\to\infty$,
an even parity ``half-bound'' threshold state enters the spectrum at
$\omega = m$  just as in the bosonic case \cite{us2}.  Also, in this
limit, the lowest (positive parity) mode approaches $\omega =
0$, and is only counted as a $\half$ as described above.
Finally, a negative parity mode enters the spectrum
from below, also to be weighted by $\half$.

\begin{figure}[htbp]
$$
\BoxedEPSF{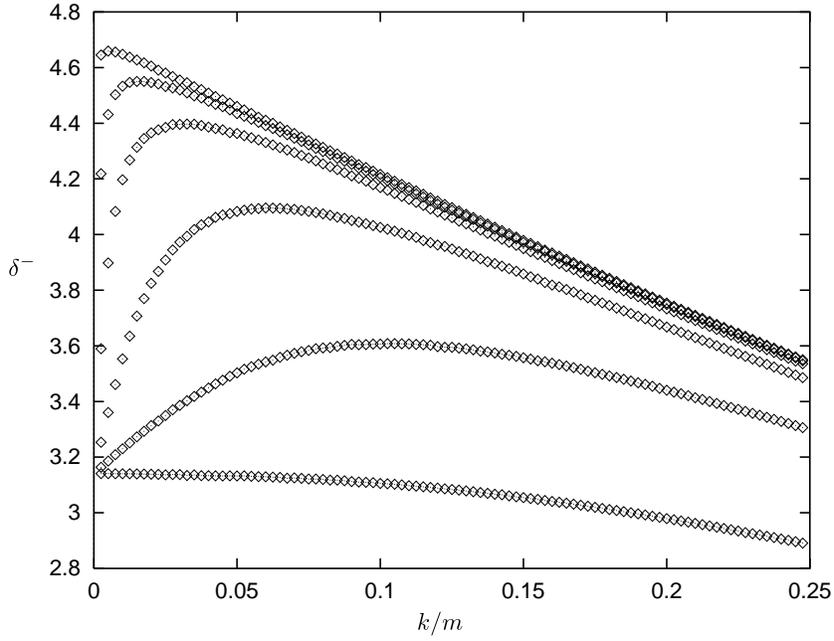 scaled 600}  
$$
\caption{Negative-parity phase shifts as functions of $k/m$ for $x_0$ =
2.0,~3.0,~4.0,~5.0,~6.0, and 8.0.  For any finite separation, the phase
shift is equal to $\pi$ at $k=0$, but as $x_0$ gets larger, the phase 
shift ascends more and more steeply to $\frac{3\pi}{2}$. }
\label{phasefa}
\end{figure}

Thus the net change is to add half of a negative parity
state, which via Levinson's theorem requires the phase shift
$\delta^{-}(0)$ to jump from $\pi$ to $\frac{3\pi}{2}$
as $x_{0}\to\infty$.  This jump occurs by the same continuous
but nonuniform process as in all cases where a new state gets bound,
which is illustrated in Fig.~\ref{phasefa}.  Just as in the bosonic
case, in the limit of infinite separation the potential we have chosen
becomes reflectionless, which requires $\delta^+(k) = \delta^-(k)$.

\begin{figure}[htbp]
$$
\BoxedEPSF{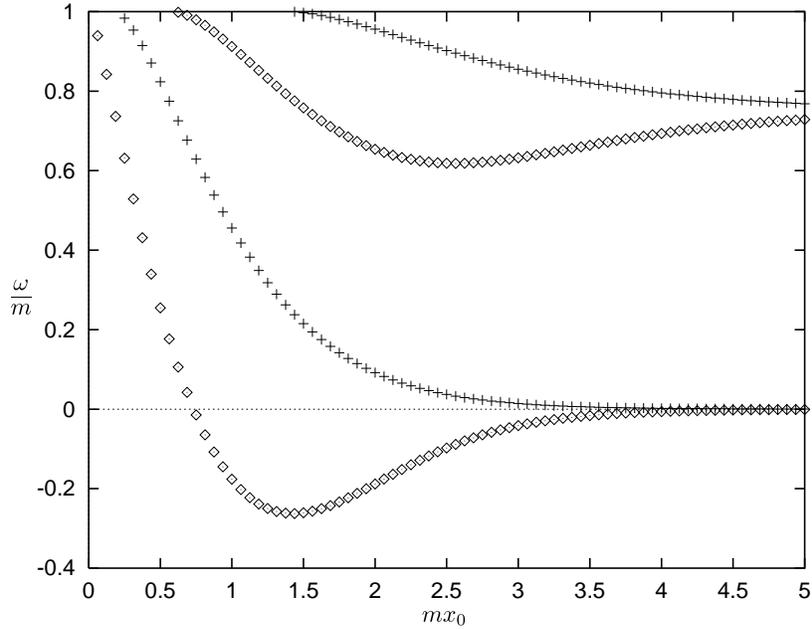 scaled 600}  
$$
\caption{Boson bound state squared energies as a function of $mx_0$ in
  units of $m$.  Symmetric states are denoted with $\diamondsuit
  \hspace{-0.525em} \cdot$~~and antisymmetric states with $+$.}
\label{boundb}
\end{figure}

To contrast the behavior of the zero modes, Fig.~\ref{boundb} shows 
$\omega^2$ for the bound states of the bosonic small oscillations 
\cite{us2}.  Because we have chosen the bosonic potential to be 
consistent with supersymmetry, the bosonic and fermionic spectra are 
related.  Again as $x_0 \to\infty$ the bound state energies approach 
those of the single soliton, and the wavefunctions are formed from the 
odd and even combinations of the wavefunctions for the single soliton.  
However, we see that in the boson case both the mode at $m 
\frac{\sqrt{3}}{2}$ and the mode at $\omega = 0$ are doubled, so there 
is no factor of $\half$ in counting the bosonic zero modes.  For $x_0$ 
above approximately $0.75/m$, the system is unstable with respect to 
small oscillations in one direction in field space.  This property is 
reflected in the lowest bound state having $\omega^2 < 0$ 
\cite{us2}.  Just as in the fermionic case, in the limit 
$x_0\to\infty$, a ``half-bound'' state enters the spectrum at $\omega 
= m$, and the phase shifts in the two channels become equal as the 
potential becomes reflectionless.

\begin{figure}[htbp]
$$
\BoxedEPSF{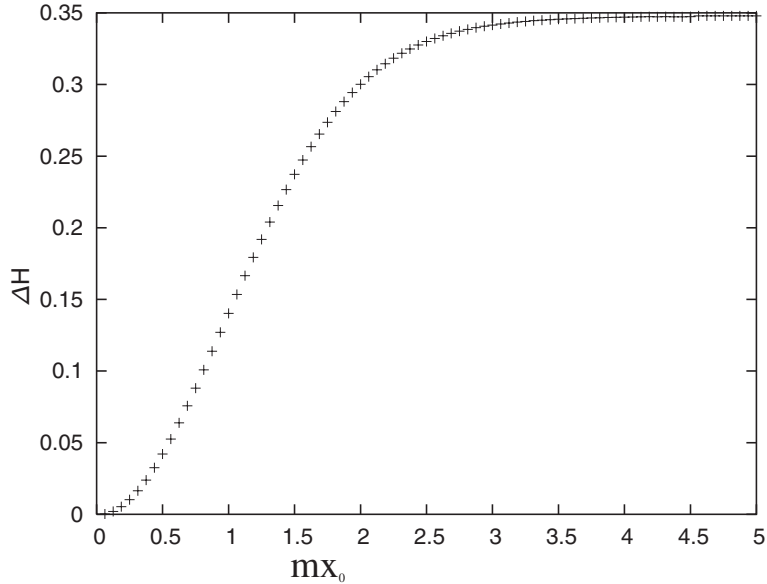 scaled 600}  
$$
\caption{One-loop fermionic correction to the energy as a function of
$x_0$ in units of $m$.}
\label{energyf}
\end{figure}
Fig.~\ref{energyf} shows the values of $\Delta H$ we obtain from
Eq.~(\ref{totenergy}) as a function of
$x_0$.  In the $x_0\to\infty$ limit, we can also do the calculation
analytically \cite{us}.  We find a phase shift
\begin{equation}
\delta_F(k) = 4\tan^{-1}\frac{m}{2k} + 2\tan^{-1}\frac{m}{k}
\end{equation}
with Born approximation
\begin{equation}
\delta_F^{(1)}(k) = \frac{4m}{k}
\end{equation}
and contributions from two bound states at $\omega =
m\frac{\sqrt{3}}{2}$ and one at $\omega = m$.  We thus find
\begin{equation}
\Delta H = 2m \left(\frac{1}{\pi} - \frac{1}{4 \sqrt{3}}\right)
\end{equation}
which agrees with the numerical calculation.  As shown in
Ref.~\cite{us2}, the bosonic contribution to the energy is
$2m \left(\frac{1}{4 \sqrt{3}} - \frac{3}{2\pi} \right)$ in this limit,
giving a total of $-\frac{m}{\pi}$ for the soliton/antisoliton
pair in the supersymmetric model, in agreement with Ref.~\cite{us}.

\section*{Discussion}

We have shown how to compute one-loop fermionic contributions to the 
energies of symmetric background field configurations.  Taking these 
methods together with those of Ref.~\cite{us2}, one can  calculate the
energy of a localized field configuration in any model  of scalars and
spinors in 1+1~dimensions.  This procedure implements  standard
renormalization conditions without ambiguity and lends itself  to robust
and efficient numerical calculation.  A natural extension of  this work is
to treat translationally invariant classical field  configurations.  Long
ago, Christ and Lee showed how to quantize the  small fluctuations about
momentum eigenstates formed by taking superpositions of {\it solutions\/}
to the classical equations of motion at different positions \cite{CL}.  New 
problems arise when studying the effects of translation invariance upon
fluctuations about time-independent classical field configurations that
are not solutions to the classical equations of motion.  Work is underway
to extend the above techniques to the nonlocal potentials encountered in
this case \cite{Sergei}.

\section*{Acknowledgments}

We wish to thank our colleagues, Eddie Farhi, Dan Freedman and Jeffrey
Goldstone for comments and suggestions on this work.  We thank Lai-Him Chan
for finding an error in our numerical calculations by comparing our results
with those of \cite{Chan}.

\section*{Appendix:  Properties of a single soliton}

As a check on the calculations presented in the body of the paper, we
present the properties (phase shifts, eigenenergies, etc.) for a
single soliton directly.  In this case, we have a potential that
satisfies $V_F(x) = -V_F(-x)$ with $V_F(x) \to\pm m$ as
$x\to\pm\infty$.  We will use the bosonic potentials that we obtain by
squaring the Dirac equation to obtain a result that agrees with the
previous method in the limit of a widely separated soliton/antisoliton
pair.  We find
\begin{equation}
  \left( \matrix{ -\frac{d^2}{dx^2} + V_B(x) & 0 \cr 0 &
      -\frac{d^2}{dx^2} + \tilde V_B(x) } \right) \psi_k(x) = k^2 \psi_k(x)
\end{equation}
where
\begin{equation}
V_B(x) = V_F(x)^2 - V_F^\prime(x) - m^2
\end{equation}
and
\begin{equation}
\tilde V_B(x) = V(x)^2 + V_F^\prime(x) - m^2
\end{equation}
are bosonic potentials.

Since $V_F(x)$ is antisymmetric, both $V_B(x)$ and $\tilde V_B(x)$ are
symmetric.  We decompose their solutions (as bosonic potentials) into
symmetric and antisymmetric channels.  For $x\to\pm\infty$, we have
\begin{eqnarray}
\eta_k^S(x) &=& \cos(kx \pm \delta^S(k)) \cr
\eta_k^A(x) &=& \sin(kx \pm \delta^A(k)) \cr
\tilde\eta_k^S(x) &=& i\cos(kx \pm \tilde\delta^S(k)) \cr
\tilde\eta_k^A(x) &=& -i\sin(kx \pm \tilde\delta^A(k))
\label{phase}
\end{eqnarray}
where the arbitrary factors of $\pm i$ are inserted for convenience
later.  By the (first-order) Dirac equation, for all $x$,
\begin{eqnarray}
  \omega_k \tilde \eta_k^S(x) &=& i\left( \frac{d}{dx} + V_F(x)
  \right) \eta_k^A(x) \cr
  \omega_k \eta_k^A(x) &=& i\left( \frac{d}{dx} - V_F(x)
  \right) \tilde\eta_k^S(x)
  \label{Dirac1}
\end{eqnarray}
and
\begin{eqnarray}
  \omega_k \tilde \eta_k^A(x) &=& i\left( \frac{d}{dx} + V_F(x)
  \right) \eta_k^S(x) \cr
  \omega_k \eta_k^S(x) &=& i\left( \frac{d}{dx} - V_F(x)
  \right) \tilde\eta_k^A(x)
  \label{Dirac2}
\end{eqnarray}
so that the solutions to the Dirac equation are
\begin{equation}
  \psi_k^+(x) = \left( \matrix{ \eta_k^S \cr \tilde \eta_k^A } \right)
\end{equation}
and
\begin{equation}
  \psi_k^-(x) = \left( \matrix{ \eta_k^A \cr \tilde \eta_k^S } \right).
\end{equation}
The phase relation between the upper and lower components of these
wavefunctions must be different as $x\to\pm\infty$ since the
mass term has opposite signs in these two limits.

Putting this all together gives, as $x\to\pm\infty$,
\begin{eqnarray}
\cos(kx \pm \delta^S(k)) &=& \frac{1}{\omega_k}
\left(\frac{d}{dx}-V_F(x)\right) \sin(kx \pm \tilde\delta^A(k)) \cr
&=& \frac{1}{\omega_k}\left(k \cos(kx \pm \tilde\delta^A(k)) \mp
m \sin(kx \pm \tilde\delta^A(k)) \right) \cr
&=& \mp \sin(kx \pm \tilde\delta^A(k) \mp \theta)  =
\cos(kx \pm \tilde\delta^A(k) \mp \theta \pm \frac{\pi}{2})
\end{eqnarray}
and thus
\begin{equation}
\delta^S(k) = \tilde\delta^A(k) + \tan^{-1} \frac{m}{k}
\end{equation}
and similarly
\begin{equation}
\delta^A(k) = \tilde\delta^S(k) + \tan^{-1} \frac{m}{k}.
\end{equation}
The fermion phase shift in each channel is given by the average of the
bosonic phase shifts of the two components
\begin{eqnarray}
\delta^+(k) &=& \half(\delta^S(k) + \tilde \delta^A(k)) =
\delta^S(k) -  \half \tan^{-1} \frac{m}{k} \nonumber\\
\delta^-(k) &=& \half(\delta^A(k) + \tilde \delta^S(k)) =
\delta^A(k) -  \half \tan^{-1} \frac{m}{k} \nonumber\\
\end{eqnarray}
so that
\begin{equation}
\delta_F(k) = \delta^+(k) + \delta^-(k) = \delta^S(k) + \delta^A(k) -
\tan^{-1}\frac{m}{k}.
\end{equation}

We note that in this case there is always a zero-energy bound state given by
\begin{equation}
  \psi(x) = \left( \matrix{ e^{-\int_0^x V_F(y) dy} \cr 0 } \right).
\end{equation}
Because it is at zero energy, this state is shared with the
antiparticle spectrum and thus only contributes with a weight of
$\half$ to Levinson's theorem and the sum over small oscillations,
as we found by following  a soliton/antisoliton pair
as a function of separation.  The extra $\tan^{-1}\frac{m}{k}$
(which goes to $\frac{\pi}{2}$ as $k\to 0$) is thus necessary to connect the
fermionic phase shift to bosonic phase shifts, which have no such
factors of $\half$ associated with their zero modes.

\end{document}